\documentstyle[12pt]{article}
\date{}
\begin{document}

\title{The Strangeness Magnetic Moment of the Proton
in the Chiral Quark Model}

\author{L. Hannelius, D.O. Riska}
\maketitle

\centerline{\it Department of Physics, University of Helsinki,
00014 Finland}

\vspace{0.5cm}

\centerline{and}

\vspace{0.5cm}

\large
\centerline{L. Ya. Glozman}

\normalsize

\vspace{1cm}

\centerline{\it Institute for Theoretical Physics, University of Graz,
A-8019 Graz, Austria}

\setcounter{page} {0}
\vspace{1cm}

\centerline{\bf Abstract}
\vspace{0.5cm}

The strangeness magnetic moment of the proton is 
shown to be small in the chiral 
quark model. The dominant loop
contribution is due to kaons. The $K^*$ loop contributions are
proportional to the difference between the strange and light constituent
quark masses or $m_{K^*}^{-2}$ and therefore small.
The loop fluctuations that involve radiative transitions 
between $K^*$ mesons and kaons are small, when the
cut-off scale in the loops is taken to be close to the
chiral symmetry restoration scale.
The net loop amplitude contribution to the strangeness magnetic
moment of the proton is about $-0.05$,
which falls within the uncertainty range of the experimental value.

\vspace{2cm}

\newpage
\centerline{\bf 1. Introduction}
\vspace{0.5cm}

	The recent finding by the SAMPLE collaboration that the
strangeness magnetic moment of the proton is small, and possibly
even positive \cite{SAMP} ($G_M^s(Q^2=0.1$ GeV$^2)=0.23\pm 0.37$)
was unexpected in view of the fact that the bulk of the many
theoretical predictions for this quantity are negative, and
outside of the experimental uncertainty range (summaries are
given e.g. in refs.~\cite{Mus94,Beis,Barz}). A recent
lattice calculation gives $-0.36\pm 0.20$ for this quantity
\cite{Dong}, thus reaffirming the typical theoretical
expectation, while remaining outside of the uncertainty
range of the empirical value.

	Under the assumption that the main strange loop 
contribution to the strangeness magnetic moment
is the kaon loop, the expectation that
$G_M^s(0)$ be negative may be explained as follows: 
Consider a $u$ quark with spin $s_z$ projection 
$+{1\over 2}$. If this fluctuates into a kaon and an $s$ quark,
the probability that $s_z$ of the $s$ quark be $-{1\over 2}$ is
twice that for the value $+{1\over 2}$ if the quarks are nonrelativistic
(for ultra-relativistic quarks the $\gamma_5$ vertex would imply 100\%
helicity flip). As the charge of the
$s$ quark is $-e/3$, the $s$ quark should then contribute a positive
amount to the net (positive) magnetic moment of the $u$ quark. 
On the other hand, following spin-flip, the $\bar s$ quark in the 
kaon will have
orbital angular momentum $l_z=+1$, and, as its
charge is $+e/3$, it should also contribute a
positive amount to the net magnetic moment of the $u$ quark. A similar
argument applies for $d$ quarks.
As the $u$ quarks contribute $4/3$ and the $d$ quarks only $-1/3$ to the
proton magnetic moment, it follows that these strange fluctuations
will give a positive contribution
to the magnetic moment of the proton as usually defined.
By convention $G_M^s$ is defined as the matrix element 
of $\bar s\gamma^\mu s$
and not of $-{e\over 3}\bar s\gamma_\mu s$, which matrix
element is obtained by 
multiplication of the ``conventional''
magnetic moment by $-3$, and thus this latter value
should be negative \cite{Mus94b}. This 
argument applies equally well at the
hadronic level, where the corresponding strange fluctuation of the 
proton is into a kaon and a strange hyperon. 

This general 
argument implies that a positive contributions to $G_M^s(0)$
can only arise from loop fluctuations that involve strange vector
mesons and loops with transition couplings between different
strange mesons. That is indeed borne out by calculations
based on the nonrelativistic harmonic oscillator
quark model \cite{Gei},
and by calculations of meson-hyperon loop fluctuations of
baryons \cite{Mus94b,Barz}.
We here show that in 
the chiral quark model, in which the strange pseudo-scalar
($K$) and vector $(K^*)$ mesons
are coupled directly to the quarks, and which admits a covariant
calculation, with coupling strengths determined by known
hadronic couplings, 
these vector meson loop contributions are much smaller
than those of the kaon loops. Consequently the sign of
the net loop contribution is determined by the kaon loops.

To explain the reason for the smallness of the diagonal
vector meson loop contribution to the strangeness magnetic
moment, it is useful to separate this into
the terms that arise from the $\delta_{\mu\nu}$
term in the vector meson spin-projection operator
$\delta_{\mu\nu}+k_\mu k_\nu/m_K^{*2}$, and 
the terms that arise from the
second $k_\mu k_\nu/m_k^{*2}$ term. The former are
proportional to the quark mass difference $m_s-m_u$, and vanish in the
$SU(3)_F$ limit $m_s=m_u=m_d$. This is due to a remarkable 
cancellation between the loop amplitudes that are associated
with 
e.m. field coupling to the intermediate
vector meson and the $s$ quark 
respectively. The latter terms are small because of the large meson mass
squared in the denominator. As a consequence vector meson
loop contributions to
the strangeness magnetic moment are small.

The present calculation should be contrasted with the
calculations in refs.~\cite{Barz,Mus94b}, wherein the 
strangeness fluctuations were considered at the hadronic
level and were found to give a large net negative result. 
The consideration of the strangeness fluctuations
of the constituent
quarks and their
contribution to the nucleon magnetic moments here
is motivated by the fact that the non-strange magnetic
moments of the nucleons are explained by
the constituent quark model. In contrast
the hadronic model is known to fail for the anomalous
magnetic moments of the nucleons, because of
the large meson-nucleon coupling constants 
and the consequent absence of a converging
loop expansion \cite{Houriet,Bethe}. The small
kaon-quark coupling constant does in the present
application imply loop
corrections, which are sufficiently small so as
not perturb the overall quark model
description of the magnetic moments \cite{GloRis99}.
Moreover in the quark model approach, which automatically
takes into account all baryonic intermediate states,
the dominant mass dependence is due
to the meson masses \cite{Gei}, with a consequent 
suppression of heavy meson loops, which is absent in the 
hadronic approach with
large baryon masses in the energy denominators.

Baryon chiral perturbation theory does not predict
the strangeness
magnetic moment, as there is no other datum, which can fix the
required strange 
counter-term in the Lagrangian density \cite{Mus2,Hemmert}.
A brief digression concerning the conceptual differences between the
treatment of mesonic fluctuations
of constituent quarks, and (directly) of baryons 
is in order.  In the latter the mesonic loop
fluctuations imply an infinite shift of the baryon mass in the chiral
limit, which is balanced by
phenomenologically determined counter-terms \cite{Gasser}. The infinite
mass renormalisation of the nucleon mass may be avoided in the heavy
baryon version of chiral perturbation theory,
as long as no decuplet states are considered
in the meson loops \cite{Mano1,Mano2}. Consideration of the mesonic
degrees of freedom in such approaches yield information on corrections
from the finite masses of the current quarks, but 
not on the origin of the nucleon mass nor on the octet-decuplet splitting
in the chiral limit. 

In the chiral constituent quark model \cite{GloRis} the role of the
meson degrees of freedom is broader. On the one hand 
meson fluctuations contribute to
the self energy of the constituent quarks, in analogy with the baryonic
approach, but on the other hand they imply a strong flavour-dependent
interactions between the quarks, which yield an octet-decuplet
splitting already in the chiral limit \cite{Glo99}. The chiral quark
model combines the $SU(6)$ flavour-spin structure of the nucleon wave
function, which is implied in the large $N_c$ limit of QCD
\cite{Dash1,Dash2}, with the dynamical implications of spontaneous
breaking of chiral symmetry. As an example both the axial current $F$
and $D$ coupling constants are predicted by the chiral quark model,
while they have to be determined by experimental data in
baryon effective field theory.

Some of the 
strange loop contributions considered here are logarithmically
divergent and thus depend on the choice of a cut-off scale. 
The natural value for this 
scale is the chiral symmetry
restoration scale $4\pi f_\pi \sim 1.2$ GeV.
With such values for the cut-off a very small value
for $G_M^s(0)$ obtains.

This paper falls into 5 sections. In section 2 the contributions
from kaonic loops are derived, while those from diagonal
$K^*$ loops are derived in section 3.
In section 4 the contribution
to the strangeness 
magnetic moments of the nucleons from the strange loops, 
which involve a $K^*K\gamma$ transition
is derived. 
Section 5 contains a concluding discussion.
We use the East coast metric throughout this paper.

\vspace{1cm}

\centerline{\bf 2. Kaon loop contributions}

\vspace{0.5cm}

Consider the (hidden) strangeness fluctuations
$q\rightarrow K s \rightarrow q$ of $u$ and $d$ quarks
that are illustrated by the Feynman diagrams in
Fig.~1 (with inclusion of seagull terms
as appropriate). The expressions for these contributions
are similar to the corresponding expressions for the 
contribution of pionic
fluctuation to the neutron magnetic moment \cite{Bethe}. For the
derivation 
of these contributions we employ the pseudo-vector 
coupling for kaons to constituent quarks:
$${\cal L}_{Kqs}=i{f_{Kqs}\over m_K}\bar\psi\gamma_\mu\gamma_5
\sum_{a=4}^7\lambda^a \partial_\mu K^a \psi,\eqno(2.1)$$
and the 
the kaon and $s$ quark current density
operators:
$$ j_\mu = ie\{\partial_\mu K^\dagger K+
\rm{h.c.}\},\eqno(2.2a)$$
$$ j_\mu = -i{e\over 3}\bar\psi_s \gamma_\mu \psi_s. \eqno(2.2b)$$
Here the underlying
assumption is that the strange quark in the loop may
be approximately described by a free fermion propagator,
with a
constant constituent mass in the relevant low momentum region. 
With the usual (unusual) convention of assigning the $s$ quark
a ``strangeness charge'' of $1$, and the kaon a ``strangeness
charge'' of $-1$, the kaon current (2.2a) should be multiplied by
$-1$ and the $s$ quark current (2.2b) by $-3$. 

The pseudo-vector kaon-quark coupling constant $f_{Kqs}$ may be related to
the kaon decay constant $f_{K}$ as
$$f_{Kqs}={g_A^q\over 2} {m_K\over f_K}.\eqno(2.3)$$
Here $g_A^q$ is the axial coupling constant for the
constituent quark, for which we use the value $g_A^q$
= 0.87 \cite{Wein,Dicus}. With $f_K=$ 113 MeV we then
obtain $f_{Kqs}=1.9$.

The pseudo-vector coupling (2.1) implies a contact current
term, which generates two seagull diagrams in addition
to the loop diagrams in Fig.~1. The contribution from these
seagull diagrams cancels the corresponding contact terms
that obtain from the diagrams in Fig.~1, once the Dirac
equation for the external quark lines is used. The remaining
term corresponds to the loop diagrams obtained if the
the kaons couple to quarks by pseudo-scalar coupling:
$${\cal L}_{Kqs}=ig_{Kqs}\psi\gamma_5\sum_{a=4}^7\lambda^a
K^a\psi, \eqno(2.5)$$
where
$$g_{Kqs}={m_q+m_s\over m_K}f_{Kqs}.\eqno(2.6)$$
Here $m_q$ and $m_s$ represent the constituent masses of
the $u,d$ and $s$ quarks respectively.
For the latter we employ the values $m_q=340$ MeV
and $m_s=$ 460 MeV \cite{GloRis}. This then yields the
value $g_{Kqs}^2/4\pi$ = 0.75 for the effective
coupling strength of the kaon loop diagrams. The smallness
of this value indicates a converging loop expansion.

While the kaon loop contributions to the strangeness
magnetic moment is finite even in the absence of
regularisation, we impose a cut-off at the chiral
restoration scale $4\pi f_\pi \sim $ 1.2 GeV. The cut-off
may be imposed by multiplying the kaon propagator 
$v(k^2)=1/(k^2+m_K^2)$ in the
loop diagram, in which the coupling is to the current of
the $s$ quark (Fig.~1a), by a squared monopole form factor:
$$v(k^2)\rightarrow
{1\over m_K^2+k^2}\left ({\Lambda^2-m_K^2\over \Lambda^2+
k^2}\right )^2.
\eqno(2.7)$$
Maintenance of the current conservation condition on the
sum of the two kaon loop contributions in Fig.~1 then
requires that the two kaon propagators in the loop amplitude,
in which the coupling is to the kaon current (Fig.~1b), be
modified in the following way \cite{Gross}:
$${1\over m_K^2+ k_1^2}{1\over m_K^2+k_2^2}\rightarrow
{v(k_2^2)-v(k_1^2)\over k_1^2-k_2^2}.\eqno(2.8)$$
These kaonic loops then give the same 
contribution to the anomalous strangeness magnetic moment of the 
$u$ and $d$ quarks. By the standard quark model result
$$\mu(p)={1\over 3}[4\mu(u)-\mu(d)],\eqno(2.9)$$
the strangeness magnetic moment contribution to the
proton from these loop diagrams is then $\mu_s(p)=
\mu_s(u)$. 
The explicit expressions for the contributions from the
two loop diagrams in Figs.~1a and b are then
$$G_M^s(0)\{a,K\}=-{g_{Kqs}^2\over 4\pi^2}{m_p\over m_q}
\int_0^1 dx(1-x)^2
{ m_q(m_s-m_q x)\over H(m_K^2)},\eqno(2.10a)$$
$$G_M^s(0)\{b,K\}=-{g_{Kqs}^2\over 4\pi^2}{m_p\over m_q}
\int_0^1 dx x(1-x)
{ m_q(m_s-m_q x)\over H(m_K^2)},\eqno(2.10.b)$$
Here the function $H(m_K^2)$ is defined as
$$H(m_K^2)=m_s^2(1-x)-m_q^2x(1-x)+m_K^2x.\eqno(2.11)$$

It follows from (2.9) that the net 
kaon loop contribution to the strangeness
magnetic moments of the proton is 
$$G_M^s(0)\{K\}=G_M^s(0)\{a,K\}+G_M^s(0)\{b,K\}.\eqno(2.12)$$

The expressions (2.9) and (2.10) do not take into account
the cut-off form factor (2.7),(2.8). The effect of the
form factor is however readily taken into account by
substitution of the factor $1/H(m_K^2)$ in (2.10) by
the replacement
$${1\over H(m_K^2)}\rightarrow K(m_K^2)\equiv
{1\over H(m_K^2)}-{1\over H(\Lambda^2)}
-x{\Lambda^2-m_K^2\over H(\Lambda^2)^2}.\eqno(2.13)$$

The values of the expression (2.12) are shown in Fig.~2 as 
functions of the cut-off parameter $\Lambda$. For 
$\Lambda$ below 1.2 GeV 
the kaon loop contribution is small and negative (as expected), and
does not exceed -0.064 $\mu_N$ in magnitude. Note the limit
$$\lim_{\Lambda\rightarrow 
\infty}\lim_{m_K\rightarrow 0}\lim_{m_s\rightarrow
m_q}G_M^s(0)\{K\}=
-{g^2_{Kqs}\over 4\pi^2}{m_p\over m_q}.\eqno(2.14)$$
If $g_{Kqs}$ is replaced by the pseudo-scalar pion-nucleon coupling
constant, and $m_q$ is replaced by the nucleon mass, this limit agrees
with the well known value for the contribution of the $\pi^-p$ loop
fluctuation to the anomalous magnetic moment of the neutron \cite{Bethe}.

\vspace{1cm}

\centerline{\bf 3. $K^*$ loop contributions}

\vspace{0.5cm}

To calculate the loop fluctuations that involve
$K^*$ mesons illustrated by the Feynman diagrams in 
Fig.~1, when the meson lines represent strange vector
mesons, we employ the Lagrangian that describes the
coupling of constituent quarks to $K^*$ mesons.

$${\cal L}_{K^*qs}=i{g_{K^*qs}\over m_K^*}\bar\psi
(\gamma_\mu
+i{\kappa_{K^*qs}\over 2\bar m}\sigma_{\mu\nu}\partial_\nu)
\sum_{a=4}^7\lambda^a K_\mu^{*a}\psi.\eqno(3.1)$$
Here $\bar m$ represents the mean constituent mass
$\bar m=(m_q+m_s)/2$.
The vector ($g_{K^* qs}$) and ratio of tensor 
to vector 
coupling constants 
($\kappa_{K^* qs}$) may be determined from the corresponding
$K^* -$ baryon octet coupling constants by 
the quark model relations \cite{Brown}:
$$g_{K^* qs}=g_{K^* \bar B B},\eqno(3.2a)$$
$$g_{K^* qs}(1+\kappa_{K^* qs})={3\over 5} 
{\bar m\over \bar M}
g_{K^* \bar B B}(1+\kappa_{K^* \bar B B}).\eqno(3.2b)$$
Here $\bar M$ represents the average of the nucleon
and $\Lambda$ hyperon masses ($\bar M= $ 1027 MeV).

The nucleon-hyperon coupling constants may be determined
indirectly by fitting potential models to nucleon-hyperon
scattering data. A recent such determination gives 
$g_{K^* \bar B B}$ = 2.97 and $\kappa_{K^* \bar B B}$
= 4.22,
although with a considerable uncertainty margin \cite{Rijk}.
From (3.2) we then have $g_{K^* qs}^2/4\pi \simeq$ 0.7
and $\kappa_{K^* qs}\simeq$ = 0.21. In view of the smallness
of the tensor coupling for the quarks we shall neglect
it here. While the vector coupling constant is large a substantial
cancellation makes the contribution from the $K^*$ loops very small, 
as shown below.

The current density operator for the $K^{*\pm}$ mesons has the
form
$$j_\mu=\pm ie\{K_\nu^{*\dagger}\partial_\mu K_\nu^*
-K_\nu^{*\dagger}\partial_\nu K_\mu^*\} + 
\rm{h.c.}\eqno(3.3)$$
The vector meson propagator is $-i/(m_K^{*2}+k^2)
(\delta_{\mu\nu}+k_\mu k_\nu/m_K^{*2})$. With the current
density operators (2.2b), (3.3) and the coupling (3.1) the
combination of the amplitudes that correspond to the
two vector meson loop diagrams in Fig.~1 
yield a conserved current.

The two vector meson loop diagrams in give the same
contributions to the anomalous strangeness magnetic
moments of the $u$ and $d$ quarks, and by (2.9) these
contributions equal that for the corresponding
strangeness magnetic moment of the proton.

The quark current loop amplitude (Fig.~1a) gives the
following contribution to $G_M^s(0)$:
$$G_M^s(0)\{a,K^*\}=2{g_{K^* qs}^2\over 4\pi^2}{m_p\over m_q}
\int_0^1 dx \Biggl\{ m_q x (1-x) [2m_s-m_q(1+x)]K(m_K^{*2})$$
$$-{1\over 2}{m_q^2\over m_K^{*2}}
(1-x)\Biggl[(1-x)^2(m_s-m_q)(m_s+m_qx)K(m_K^{*2})$$
$$-2\left(1-{m_s\over m_q}\right)
\left(1-{3\over 2}x\right)
\left (\ln {H(\Lambda^2)\over H(m_K^{*2})}-x{\Lambda^2
-m_K^{*2}\over H(\Lambda^2)}\right )\Biggr]\Biggr\}.\eqno(3.4)$$
The function $K(m_K^{*2})$ is defined in (2.13).
A check of this expression is possible by comparison
to the anomalous magnetic moment of the electron. Consider
the limit
$$\lim_{\Lambda\rightarrow \infty}
\lim_{m_K*\rightarrow0}\lim_{m_s\rightarrow m_q}G_M^s(0)\{a,K^*\}=
{g_{K^* qs}^2\over 4\pi^2}{m_p\over m_q}.\eqno(3.5)$$
For comparison to the anomalous magnetic moment of
the electron divide out the flavour factor 2 
and change the sign to make allowance for the 
negative electron
charge. The result then reduces to the standard 
value $-\alpha/2\pi$ for the anomalous magnetic moment
of the electron if the quark and proton masses are
replaced by the electron mass, and $\alpha$ is
substituted for $g^2_{K^* qs}/4\pi$.

The contribution from the $K^*$ current loop amplitude (Fig.~1b) 
to $G_M^s(0)$ is
$$G_M^s(0)\{b,K^{*}\}= -{g_{K^* qs}^2\over 4\pi}
{m_p\over m_q}\int_0^1dx x\Biggl\{m_q
[m_s(2-3x)-m_q x (1-2x)]K(m_K^{*2})$$
$$-{m_q^2\over m_K^{*2}}\Biggl [m_q(1-x)^2(m_s-m_qx^2)
K(m_K^{*2})$$
$$-2\left({m_s\over m_q}-2(1-x)^2-x\right)
\left (\ln{H(\Lambda^2)\over H(m_K^{*2})}-x{\Lambda^2-m_K^{*2}
\over H(\Lambda^2)}\right )\Biggr ]\Biggr\}.\eqno(3.6)$$

The net $K^*$ loop contribution to the strangeness
magnetic moment of the proton is then [cf. (2.12)]
$$G_M^s(0)\{K^*\}=G_M^s(0)\{a,K^*\}+G_M^s(0)\{b,K^*\}.\eqno(3.7)$$
This contribution to the strangeness magnetic moment of
the proton is also shown in Fig.~2.
The vector meson loop contributions is much smaller in magnitude than
the kaon loop contribution. It changes sign
at $\Lambda \sim$ 1.2 GeV.
The smallness of the vector meson
loop contribution may also be understood from the
fact that the sum of the main terms in (3.4) and (3.5)
(the terms without the factor $1/m_{K^*}^2$) is
proportional to the (small) quark mass difference 
$m_s-m_q$, and vanishes in the equal mass limit for the
constituent quarks. The other terms are small, as they are proportional
to $1/m_K^{*2}$. There is no {\it a priori} reason to expect the 
vector meson
loop contributions to have any specific sign, as both the spin non-flip
charge as well as the spin-flip current couplings contribute 
to the net result. In the ultra-relativistic limit, only the 
helicity conserving charge coupling survives, and with a
consequent positive loop contribution.

The cancellation of the main vector meson loop contributions to the
strangeness magnetic moment in the equal quark mass limit is remarkable,
and also applies for loop contributions considered at the hadronic
level in the case of pure Dirac couplings.
The large $K^*$ loop contributions found in ref.~\cite{Barz} 
can therefore only be understood as consequences of
large tensor couplings.

\vspace{1cm}
\newpage

\centerline{\bf 4. Strangeness Loops with Transition Couplings}

\vspace{0.5cm}

We finally consider the (hidden) strangeness fluctuations of light $u$ and 
$d$ quarks, which involve the radiative transition coupling
$K^*\rightarrow K\gamma$ (Fig.~3). The key vertex in this loop diagram
is the $K^*\rightarrow K\gamma$ vertex, which is described
by the transverse current matrix element:
$$\langle K^a(k')|J_\mu|K_\sigma^{*b}(k)\rangle =
-i{g_{K^*K\gamma}\over m_K^*}
\epsilon_{\mu\lambda\nu\sigma} k_\lambda k_\nu^{'} 
\delta^{ab}.\eqno (4.1) $$
The coupling constant $g_{K^*K\gamma}$ is determined
by the empirical decay widths for radiative decay of the 
$K^*$ as
$$\Gamma(K^*\rightarrow K\gamma)=\alpha{g^2_{K^*K\gamma}\over 24}
m_K^*\left [1-\left ({m_K\over m_K^*}\right )^2
\right ]^3, \eqno (4.2)$$
where $\alpha$ is the fine structure constant and $m_K$
is the mass of the kaon. From the
empirical decay widths $\Gamma(K^{*+}\rightarrow K^+\gamma)=50$ keV
and $\Gamma(K^{*0}\rightarrow K^0\gamma)=116$ keV 
\cite{PDG} we obtain
$g_{K^{*+}K^+\gamma}=0.75$ and $g_{K^{*0}K^0\gamma}=1.14$. 
 
As the decay widths do not determine the signs of these coupling
constants, we fix those by invoking $SU(3)$ flavour symmetry to
link the vertices (2.1) to the corresponding vertex for radiative
decay of the $\rho$ meson. The sign of the $g_{\rho\pi\gamma}$
coupling is ultimately determined by the triangle anomaly,
or in the Skyrme model by the Wess-Zumino term through the
topological baryon current and determines the sign of the
$\rho\pi\gamma$ exchange current contribution to
the iso-scalar exchange current of nuclei \cite{Nym,Weis}. 
With this justification, the signs above obtain.

As the $K^*Ks$ loop diagrams are logarithmically divergent, a
cut-off has to be invoked in their evaluation. Such is introduced
by insertion of monopole form factors of the form
$(\Lambda^2-m_K^2)/(\Lambda^2+k_1^{2})$ and
$(\Lambda^{*2}-m_K^{*2})/(\Lambda^{2*}+k_2^{2})$ at the $K$ and $K^*$
vertices respectively. The cut-off mass $\Lambda^*$ is taken 
equal to that of $\Lambda$ in the numerical calculations here.

Given the current matrix element (4.1) and the
Lagrangian densities (2.5), (3.1a) the contributions to the
anomalous magnetic moment of the $u$ and $d$ quarks from the
$K^*K\gamma$ loop diagrams (Fig.~3) take the following
form (when expressed in terms of nuclear magnetons and after
assigning the kaon line a ``strangeness charge'' of $-1$):
$$G_M^s(0)_{u,d}=-{g_{Kqs}g_{K^*qs}g_{K^*K\gamma}\over
2\pi^2} {m_p\over m_K^*}
\int_0^1 dx x$$
$$\times \int_0^1dy\Biggl\{ m_q(1-x)(m_s-m_q x)
\left ({1\over G_1}-{1\over G_2}-{1\over G_3}+{1\over G_4}\right )
-\ln\left ({G_2 G_3\over G_1 G_4}\right )\Biggr\} .\eqno(4.3)$$
Here $m_p$ is the proton mass and $m_s$ the constituent mass
of the $s$ quark, which we shall take to have the value
460 MeV \cite{GloRis}. The quantities $G_j$ above are
defined as
$$G_1=G(m_K,m_{K^*}),\, G_2=G(\Lambda,m_{K^*}),\,
G_3=G(m_K,\Lambda^*),\, G_4=G(\Lambda,\Lambda^*), \eqno(4.4)$$
where the function $G$ is defined as
$$G(m,m')=m_s^2(1-x)-m_q^2x(1-x)+m^2x(1-y)+m^{'2}xy.\eqno(4.5)$$
In the case of the $u$ quark the coupling $g_{K^*K\gamma}$
constant is that for the $K^+,K^{*+}$
mesons and in the case of the $d$ quark it is that for the
$K^0,K^{*0}$ mesons.

The contribution from these loops to the strangeness magnetic
moment of the proton is obtained as
$$G_M^s(0)\{K,K^*\}={4\over 3}G_M^s(0)_u\{K,K^*\}-
{1\over 3}G_M^s(0)_d\{K,K^*\},\eqno (4.6)$$
and thus will have a value close to the corresponding
constituent quark values.

In Fig.~2 we show the contribution from the $K^* Ks$ loop
to the strangeness magnetic moment of the $u$ and $d$
quarks as functions of the cut-off parameter $\Lambda$.
For $\Lambda$ values in the range 1--1.5 GeV the 
contributions are positive, but small (Table 1).

	These values are somewhat smaller than the
corresponding value obtained in ref.~\cite{Gei} with
the harmonic oscillator quark model. The difference is
mainly due to the fact that the latter is non-covariant,
and consequently overestimates the radiative widths
of the vector mesons. With the usual static fermion 
currents of the constituent quark model the coupling
constant $g_{K^*K\gamma}$ should be 2, which is about
twice too large. Even so it best to view the results in
Fig.~2 as representing lower estimates from of the
strange fluctuations, which involve radiative transition
couplings of the form $K^*\rightarrow K\gamma$, as the
higher lying vector mesons $K^*(1410)$ should also
contribute. As the radiative widths of these vector excited
strange vector mesons are not known, their contribution
cannot be calculated without an explicit quark model
at this time. If their radiative widths should be as large as
the empirical radiative widths of
the strange tensor mesons $K_2^*(1430)$ the curves in Fig.~2
would lie about 30 \% higher than those shown.  

\vspace{1cm}

\centerline{\bf 5. Discussion}

\vspace{0.5cm}

The main result obtained here is that the strangeness 
magnetic moment
of the proton is small. When the cut-off parameter has values
close to the chiral symmetry restoration scale, the net
value for it is about $-0.05$ n.m.
The smallness of the (dominant) kaon fluctuations is a
consequence of the small effective K-quark coupling constant.
The diagonal strange vector meson loop fluctuations are small
because of a cancellation between the two loop diagrams
in Fig.~1, when the meson lines represent $K^*$ mesons.  
In the $SU(3)_F$ limit the
vector meson loop contributions cancel exactly, with exception
for the terms that
arise from the $k_\mu k_\nu/m_{K^*}^2$ terms in the 
vector meson propagator, which terms are small because of the
large mass of the $K^*$. The strangeness fluctuations
that involve heavier strange mesons will be similarly
suppressed by the larger meson masses. A systematic treatment
of such will have to await the empirical determination of the
radiative widths for $K\gamma$ decay of all strange mesons
with masses below 1.5 GeV (At this time only the radiative
widths of the tensor mesons $K_2^*(1430)$ in this  region
are known). In principle
the quark model calculation of ref.~\cite{Gei} takes into account
all such effects, although it is schematic in being
nonrelativistic and in its description of the $K$ mesons
as pure quark-antiquark states rather than as the collective
states implied by their role as Goldstone bosons of the
spontaneously broken approximate chiral symmetry. The nonrelativistic
description of the $K^*K\gamma$ vertex leads to an overestimate
of the associated loop fluctuation.

The present approach recommends itself over the hadronic
loop calculations in refs.~\cite{Barz,Mus94b} by the fact
that it appears to account reasonably well also for the non-strange
baryon magnetic moments (once the appropriate exchange current 
contributions are included \cite{Dann}) and by the fact that
the chiral quark model, with the meson quark couplings
used above, also describes the baryon spectrum itself
fairly well \cite{Pless}. While in baryon effective field theory
the anomalous parts of the baryon magnetic moments are
exclusively due to loop contributions, the bulk of these are already
implied by the $SU(6)_{FS}$  structure of the wave functions in the
quark model, leaving room for only very small loop contributions
\cite{GloRis99}. That this should be so is natural as the loop
contribution to the anomalous magnetic moments at the quark level should
be suppressed by $(m_q/\Lambda_\chi)^2\sim 0.1$, where
$\Lambda_\chi=4\pi f_\pi$ is the chiral restoration scale.  

The small magnitude of the net predicted strangeness magnetic moment
$\sim -0.05$ is fairly insensitive to the value of
the cut-off parameter $\Lambda$. It falls within the uncertainty 
range of the
empirical value, and is also commensurate with recent more
phenomenological results \cite{Speth}. 
The result suggests 
that the strangeness radius of the proton
should be small, if it - in analogy with the result the neutron charge
radius \cite{GloRis99,Alex} - it is dominated by the second (Foldy) term in the
expression
$$\langle r_s^2\rangle =-6 \lim_{q^2\rightarrow 0} {d\over dq^2}F_{1s}(q^2)+
{3\over 2}{F_{2s}(0)
\over m_n^2}.\eqno (5.1)$$
Since $F_{2s}(0)=G_M^s(0)$ is small, only a remarkably strong momentum
dependence in the strange Dirac form factor 
$F_{1s}$ would lead to a substantial
value for the strangeness radius.

The only predictions for $G_M^s$, which are large and positive
(0.37, 0.41) have been obtained in the chiral bag model 
\cite{Hong,Hong1}
and a linear quark-soliton model \cite{Goek} , both of which treat
strangeness as a collective
zero mode, with large phenomenological flavour symmetry  breaking terms.  
There is no obvious way
to connect these results
to those obtained by more conventional phenomenological models.
The bound state version of the topological soliton
model \cite{Call}, in which the hyperons
are described as bound states of kaons and topological
solitons, would give the same (negative) sign for the strangeness 
magnetic moment as the conventional hadronic loop calculation.

\vspace{1cm}
\centerline {\bf Acknowledgements}

\vspace{0.5cm}

This
work was supported in part by the Academy of Finland under 
contract 43982. L. H. thanks the Finnish Society of Sciences
and Letters for support. 
D. O. R. thanks Professor R. D. McKeown 
for instructive discussions. 

\newpage

\centerline{\bf Table 1}
\vspace{0.5cm}

 Contributions to the strangeness magnetic moment of the proton
(in nuclear magnetons) 
from the loops with $K^*\rightarrow K\gamma$ transition
vertices (Fig.~1) and with diagonal $KK\gamma$ vertices
(Fig.~3) as a function of the cut-off parameter
$\Lambda$ (in MeV). The chiral symmetry restoration scale
corresponds to $\Lambda_\chi \sim 1.2$ GeV

\vspace{1cm}

\begin{center}
\begin{tabular}{|l|l|l|l|l|} 
\hline
&&&&\\
$\Lambda$ &$KK\gamma$ & $K^*K^*$ &$K^*K\gamma$ & Sum\\
&&&&\\ \hline
&&&&\\
700&-0.014&0.003&-0.005 &-0.015\\
&&&&\\
800&-0.024&0.0005&-0.003&-0.027\\
&&&&\\
900&-0.035&0.000002&0.0004&-0.035\\
&&&&\\
1000&-0.045&0.0002&0.005&-0.040\\
&&&&\\
1100&-0.055&0.0003&0.011&-0.043\\
&&&&\\
1200&-0.064&0.00008&0.018&-0.046\\
&&&&\\
1300&-0.072&-0.0005&0.025&-0.048\\
&&&&\\
1400&-0.080&-0.001&0.032&-0.049\\
&&&&\\
1500&-0.087&-0.003&0.040&-0.049\\
&&&& \\
\hline
\end{tabular}
\end{center}

\newpage
{\bf Figure Captions}
\vspace{0.5cm}

Fig.~1 The $Ks$ and $K^*s$ loop fluctuations with diagonal
couplings, which contribute to the strangeness
magnetic moments of the $u$ and $s$ quarks and
to $G_M^s(0)$ of the proton. In the case of the $K^*$ loops
the meson lines represents $K^*$ mesons.

Fig.~2 The contribution to $G_M^s(0)$ of the proton:
$K$ loops (``Kaon''), $K^*$ loops (``$K^*$''), 
$K^* K$ loops (``$K^* -K$'')
The net result is given by the curve ``Total''.

Fig.~3 The $K^* Ks$ fluctuations of $u$ and $d$ quarks, 
which involve
the radiative transition coupling $K^*\rightarrow K\gamma$.


\vspace{1cm}

\centerline{\bf References}
\vspace{0.5cm}
\begin {enumerate}

\bibitem {SAMP} B. Mueller {\it et al.}, Phys. Rev. Lett. {\bf 78} (1997) 3824
\bibitem {Mus94} M. J. Musolf {\it et al.}, Phys. Rept. {\bf 239} (1994) 1
\bibitem {Beis} E. J. Beise {\it et al.}, Proc. SPIN96 symposium, eprint
nucl-ex/9610011
\bibitem {Barz} L. L. Barz {\it et al.}, Nucl. Phys. {\bf A640} (1998)
259
\bibitem {Dong} S. J. Dong, K. F. Liu and A. G. Williams,
eprint hep-ph/9712483
\bibitem {Mus94b} M. J. Musolf and M. Burkhardt, Z. Phys. {\bf C61}
(1994) 433
\bibitem {Gei} P. Geiger and N. Isgur, Phys. Rev. {\bf 55} (1997) 299
\bibitem{Houriet} A. Houriet, Helv. Phys. Acta {\bf 18}
(1945) 473
\bibitem {Bethe} H. A. Bethe and F. de Hoffman, {\it Mesons and
Fields, Vol. II}  Row, Peterson, Evanston (1955)
\bibitem {GloRis99} L. Ya. Glozman and D. O. Riska,
Physics Letters {\bf B459} (1999) 49
\bibitem {Mus2} M. J. Ramsay-Musolf and H. Ito, Phys. Rev. {\bf C55}
(1997) 3066
\bibitem {Hemmert} T. R. Hemmert, U.-G. Meissner and S. Steininger,
Phys. Lett. {\bf B437} (1998) 184
\bibitem {Gasser} J. Gasser, M. E. Sainio and A. Svarc, Nucl. Phys. {\bf
B307} (1988) 779
\bibitem {Mano1} E. Jenkins and A. Manohar, in 
{\it Effective field theories of the standard model}, 
U.-G. Meissner ed., World Scientific, Singapore (1992) 113
\bibitem {Mano2} E. Jenkins and A. Manohar, Phys. Lett. {\bf B255}
(1991) 558
\bibitem {GloRis} L. Ya. Glozman and D. O. Riska, Phys. Rept. {\bf 268}
(1996) 263
\bibitem {Glo99} L. Ya. Glozman, Physics Letters {\bf B459} (1999) 589
\bibitem {Dash1} R. F. Dashen, E. Jenkins and A. Manohar, Phys. Rev.
{\bf D49} (1994) 4713
\bibitem {Dash2} R. F. Dashen, E. Jenkins and A. Manohar, Phys. Rev.
{\bf D51} (1995) 3697
\bibitem{Wein} S. Weinberg, Phys. Rev. lett. {\bf 65} (1990) 1181
\bibitem{Dicus} D. A. Dicus {\it et al.}, Phys. Lett.    {\bf B284}
(1992) 384
\bibitem{Gross} F. Gross and D. O. Riska, Phys. Rev. {\bf C36} 
(1987) 1228
\bibitem{Brown} D. O. Riska and G. E. Brown, hep-ph/9902319
\bibitem{Rijk} T. A. Rijken and V. G. J. Stoks,
Phys. Rev. {\bf C59} (1999) 21
\bibitem {PDG} C. Caso {\it et al.}, The European Physical Journal {\bf C3}
(1998) 1
\bibitem {Nym} E. M. Nyman and D. O. Riska, Phys. Rev. Lett.
{\bf 57} (1986) 3007, Nucl. Phys. {\bf A468} (1987) 473
\bibitem {Weis} M. Wakamatsu and W. Weise, Nucl. Phys. {\bf A477}
(1988) 559
\bibitem {Dann} K. Dannbom {\it et al.}, Nucl. Phys. {\bf A616} (1997) 555
\bibitem{Pless} L. Ya. Glozman {\it et al.}, Phys. Rev. {\bf D58} 
(1998) 094030
\bibitem{Alex} Yu. A. Aleksandrov, Phys. Part. and Nucl.
{\bf 30} (1999) 29
\bibitem{Speth} U. -G. Meissner, V. Mull and J. Speth,
Phys. Lett.{\bf B408} (1997) 381
\bibitem {Hong} S.-T. Hong, B.-Y. Park, Nucl. Phys. {\bf A561} (1993) 525
\bibitem {Hong1} S.T. Hong, B.-Y. Park and D. P. Min, 
Phys. Lett. {\bf B414} (1997) 229
\bibitem {Goek} H.-C. Kim {\it et al.}, Phys. Rev. {\bf D58} (1998) 114027
\bibitem {Call} C. G. Callan, K. Hornbostel and I. Klebanov,
Phys. Lett. {\bf B202} (1988) 269

\end{enumerate}
\end {document}
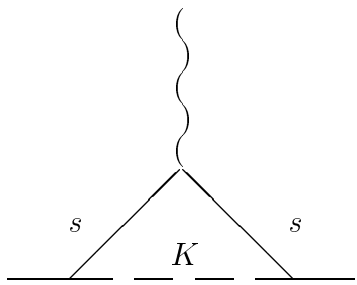
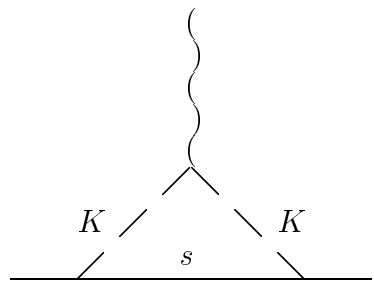
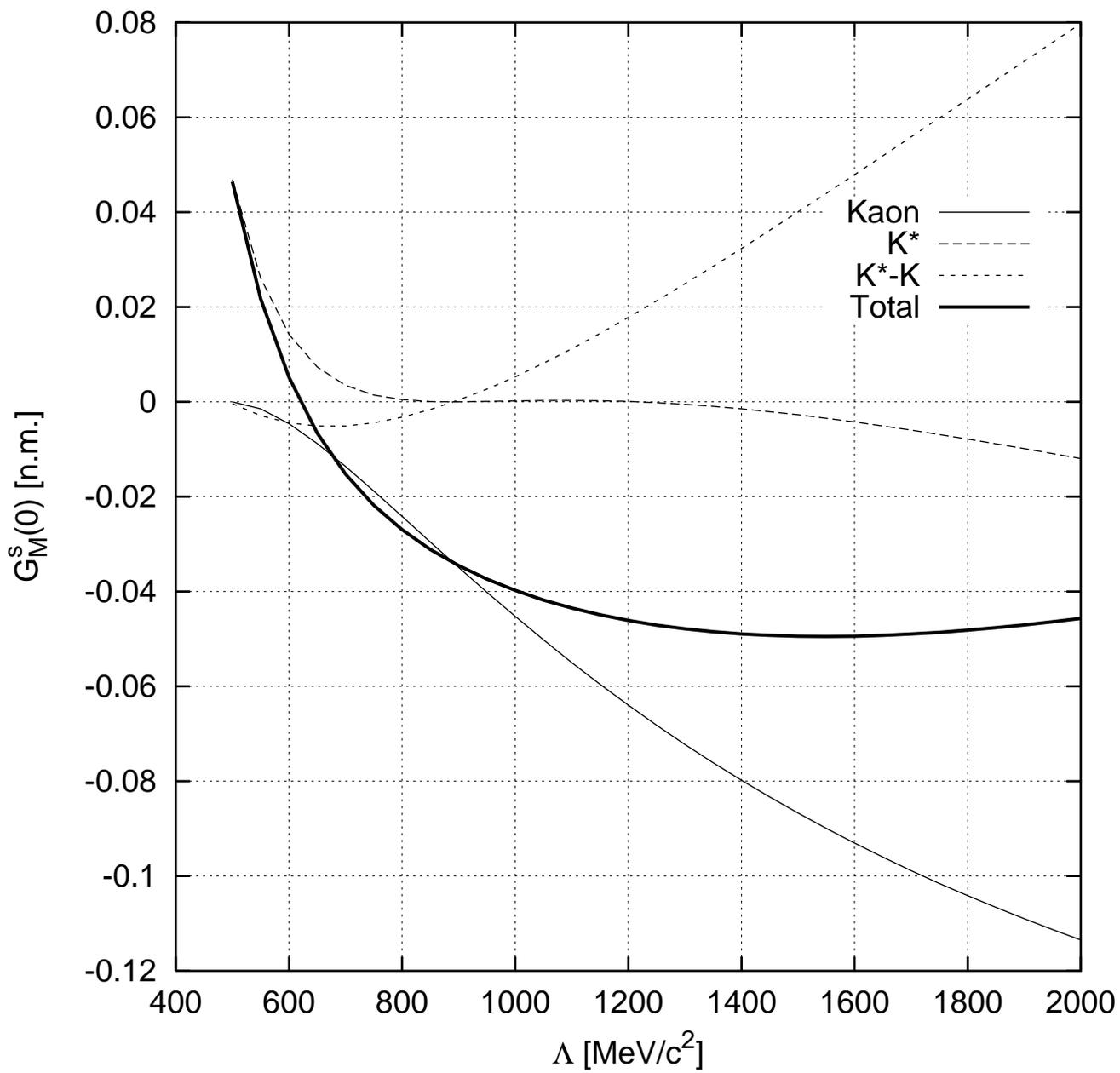
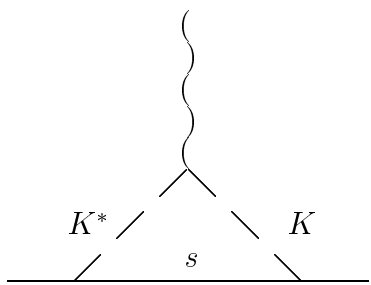
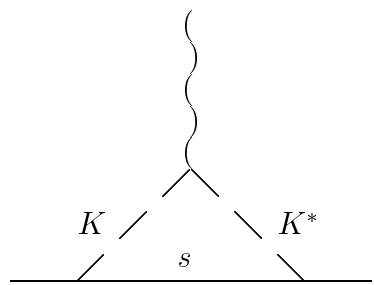
